\documentclass{emulateapj}

\usepackage{amssymb}

\slugcomment{ApJSup accepted (Spitzer Special Issue)}

\shorttitle{Spitzer/IRS observations of SBS\,0335-052}
\shortauthors{Houck et al.}

\begin{document}


\title{The Extraordinary Mid-infrared Spectrum of the Blue Compact
Dwarf Galaxy SBS\,0335-052}

\author{J. R. Houck\altaffilmark{1}, V.
  Charmandaris\altaffilmark{1,2}, B. R. Brandl\altaffilmark{3,1}, D.
  Weedman\altaffilmark{1}, T. Herter\altaffilmark{1}, L.
  Armus\altaffilmark{4}, B. T.  Soifer\altaffilmark{4}, J.
  Bernard-Salas\altaffilmark{1}, H. W. W.  Spoon\altaffilmark{1}, D.
  Devost\altaffilmark{1}, K. I.  Uchida\altaffilmark{1}}

\altaffiltext{1}{Astronomy Department, Cornell University, Ithaca, NY 14853}

\altaffiltext{2}{Chercheur Associ\'e, Observatoire de Paris, F-75014, Paris, France}

\altaffiltext{3}{Leiden University, 2300 RA Leiden, The Netherlands}

\altaffiltext{4}{Spitzer Science Center, California Institute of
  Technology, 220-6, Pasadena, CA 91125}

\email{jrh13@cornell.edu}


\begin{abstract}
  
  SBS\,0335-052 is a blue compact dwarf galaxy (BCD) with one of the
  lowest known metallicities, Z$\sim$Z$_{\sun}$/41, making it a local
  example of how primordial starburst galaxies and their precursors
  might appear.  A spectrum obtained with the Infrared Spectrograph
  (IRS\footnote{The IRS was a collaborative venture between Cornell
    University and Ball Aerospace Corporation funded by NASA through
    the Jet Propulsion Laboratory and the Ames Research Center.})  on
  the Spitzer Space Telescope clearly shows silicate absorption
  features, emission lines of [SIV] and [NeIII], and puts strong upper
  limits on the PAH emission features.  The observed low resolution
  spectrum (R$\sim$90) extends from 5.3 to 35\,$\mu$m and peaks at
  $\sim$28\,$\mu$m.  The spectrum is compared to IRS observations of
  the prototypical starburst nucleus NGC\,7714.  SBS\,0335-052 is
  quite unlike normal starburst galaxies, which show strong PAH bands,
  low ionization emission lines, and a continuum peak near 80\,$\mu$m.
  The continuum difference for $\lambda >30\,\mu$m implies a
  substantial reduction in the mass of cold dust.  If the spectrum of
  this very low metallicity galaxy is representative of star forming
  galaxies at higher redshifts, it may be difficult to distinguish
  them from AGNs which also show relatively featureless flat spectra
  in the mid-IR.

\end{abstract}


\section{Introduction}

By virtue of their low metallicities, small size and low optical
luminosity (M$_{B}\geq$-18), Blue Compact Dwarf galaxies (BCDs) may be
local examples of the building-blocks for the earliest galaxies
\citep{Rees98}. A large number of BCDs have been identified based on
their blue colors and very low metallicities as determined by optical
spectroscopy, which implies that star formation has begun only
recently \citep[see][ for a review]{Kunth00}. Their blue color arises
from one or more intense regions of active star formation that often
appear nearly devoid of dust.  The first identified member of this
class was I\,Zw\,18 \citep{Searle72}, which still remains the most
metal poor member of the class with Z=Z$_{\sun}/50$.  Another object
with similarly low metallicity Z=Z$_{\sun}$/41, is SBS\,0335-052
\citep{Izotov97} at a distance\footnote{Throughout this paper we
  assume a flat $\Lambda$ dominated universe with
  H$_0$=71kms$^{-1}$Mpc$^{-1}$, $\Omega_{\rm M}$=0.27, $\Omega_{\rm
    \Lambda}$=0.73.}  of 57.6\,Mpc. In the case of SBS\,0335-052 six
regions of massive star formation, five visible and one obscured, have
been identified, with ages less than 25Myr and a total luminosity of
$\sim10^9$L$_{\sun}$.  All six are within a region of $\sim$2$''$ or
$\sim$500pc in size \citep{Thuan97}, but roughly 75\% of the total
luminosity comes as mid-infrared (mid-IR) radiation \citep{Plante02}!
A few similar cases of obscured super star clusters which contribute a
considerable fraction of the total luminosity of the host galaxy in
the mid-IR have been observed so far. These include the Antennae
interacting galaxies \citep{Mirabel98} and the Wolf-Rayet galaxy
He\,2-10 \citep{Vacca02}.  The frequency and implications of this
phenomenon at higher redshifts (z$>$0.1) will require higher
sensitivity observations of the mid-IR emission than have been
available to date.  SBS\,0335-052 was known to have an
infrared-emitting dust continuum \citep{Thuan99}, and its dust
properties have already been studied quite extensively with both the
Infrared Space Observatory (ISO) and from ground based telescopes
\citep{Dale01,Plante02}.  However, there is still disagreement
regarding whether the emitting dust is optically thick \citep{Thuan99}
or optically thin \citep{Dale01}.  This impacts the question of how
enshrouded the first generations of stars can be and whether or not
there may exist a population of ``optically-quiet'' objects which are
visible in the infrared but not in the optical.

We chose SBS\,0335-052, along with approximately 20 other well studied
BCDs, to investigate the characteristics of star formation at very low
metallicities using the IRS \citep{Houck04} on Spitzer
\citep{Werner04}.  Similar low metallicity objects may be detected at
much higher luminosity and much greater redshifts in Spitzer discovery
surveys.

\section{Observations}

SBS\,0335-052 was observed using both IRS low resolution modules. The
spectrum extends from 5.3 to 35\,$\mu$m and was obtained on 6 February
2004. The red peak-up camera in medium-accuracy mode was used to
locate the mid-IR centroid of the source and move it to the center of
the spectrograph slits.  The total integration time was 42 minutes
with 28 minutes for the Short-Low module and 14 minutes for the
Long-Low module.  The total elapsed time including the telescope slew,
target acquisition, settling, array conditioning, and integration was
61.3 minutes.

The basic processing of the data, such as ramp fitting, dark sky
subtraction, removal of cosmic rays, droop and linearity correction,
wavelength calibration, etc, was performed using the IRS pipeline at
the Spitzer Science Center (version S9.1). The resulting spectral
images were sky-subtracted and a one-dimensional spectrum then
extracted.  The peak-up images were also used to derive a photometric
point at 22\,$\mu$m (filter bandwidth 18.5--26.0\,$\mu$m). As described in
detail in chapter 7 of the Spitzer Observer's
Manual\footnote{http://ssc.spitzer.caltech.edu/documents/som/}, during
IRS peak up we obtain 6 images of the science target. The on source
time for the 22\,$\mu$m peak-up images of our target was 6$\times$8=48
seconds and each image was created by reading the array in
double-correlated sample mode. We processed the data on the ground to
remove cosmic rays and the residual noise of the electronics. The
resulting image had a prominent diffraction ring and was
indistinguishable from the image of a point source. The conversion to
flux density was based on a number of calibration stars for which
peak-up images, IRS spectra, and reliable templates are available
\citep{Cohen03}. We find that the 22\,$\mu$m flux density of
SBS\,0335-052 is 70$\pm$11\,mJy.

\section{Results}

\subsection{Mid-IR Spectral Properties}

Figure 1 shows the 5.3--35\,$\mu$m spectrum of SBS\,0335-052 as
observed by the IRS. Our data are in good agreement with the overall
shape and intensity of the 5--15\,$\mu$m ISO spectrum of
\citet{Thuan99}, but our signal to noise is at least a factor of 10
higher. This enables us for the first time to directly detect a few
mid-IR ionic lines, while placing strong upper limits on others. The
9.7\,$\mu$m silicate absorption feature is clearly evident in our
spectrum, and the 18\,$\mu$m feature is probably present. Using the
silicate absorption profile measured towards the Galactic center
\citep{Chiar04} we find that for a screen model A$_{9.7\mu \rm m}$=
0.49\,mag, assuming a blackbody background source. More importantly,
our spectrum indicates that the emission from the galaxy peaks at
$\sim$28\,$\mu$m which, as we discuss in the following section, has
important consequences in estimating the dust mass and grain size
distribution.  In Figure 1 we also present a scaled version of the IRS
low-resolution spectrum for the prototype starburst nucleus of
NGC\,7714 \citep{Brandl04}, normalizing its flux to the corresponding
flux of SBS\,0335-052 at 14\,$\mu$m (the actual flux for NGC\,7714 is
$\sim$9.5 times larger than shown).  As also noted by \citet{Thuan99},
a striking difference between the spectrum of SBS\,0335-052 and that
of a more typical starburst is the absence of strong polycyclic
aromatic hydrocarbon (PAH) features and low excitation ionic lines. In
starburst galaxies, emission from PAHs is thought to originate from
the photodissociation envelopes bordering the \ion{H}{2} regions
produced by the ionizing starburst.  How these features can be absent
in a low-metallicity starburst is an important astrophysical question.
One possibility is that the absence of PAHs is due to low abundance of
carbon and/or nucleating grains; another possibility is that PAHs are quickly
destroyed.  Table 1 shows that the observed line ratios of
log([SIV]/[SIII])$\geq$0.48, and log([NeIII]/[NeII])$\geq$0.69 are
similar to the most extreme example of ultra-compact \ion{H}{2}
regions in our Galaxy.  The ratios indicate that the radiation field
is extremely hard and corresponds to an effective stellar temperature
of T$_{eff}\geq4\times10^4$K, assuming solar abundance
\citep[see][]{Martin02}. This would suggest that the absence of PAHs
results from their destruction by the hard UV photons and strong winds
produced by the massive stars \citep[i.e.][]{Allain96}.  Such a
scenario is quite likely since in low metallicity systems the
attenuation of UV photons is small and consequently their mean free
path in the interstellar medium can be larger and photodissociation of
PAH may occur over considerably larger scales than those seen in
typical Galactic \ion{H}{2} regions.  Whatever the PAH life-cycle is
in SBS\,0335-052, the primary result is the well-defined absence of
these features in the mid-IR.

\begin{figure}
\figurenum{1}
\epsscale{1.2}
\plotone{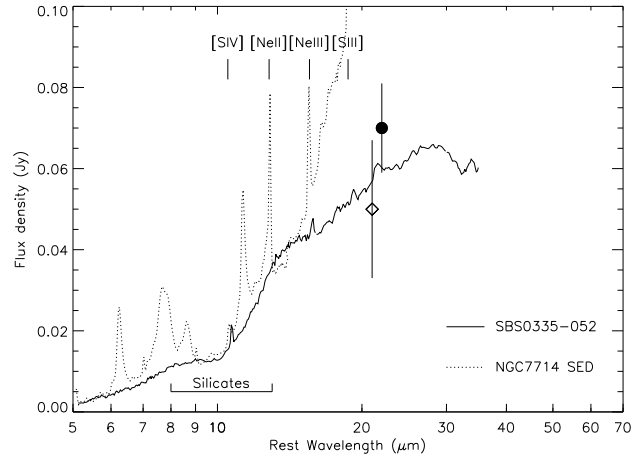}
 
\caption{
  The IRS low resolution spectrum of SBS\,0335-052 is presented (solid
  line) along with the spectrum of NGC\,7714 (dotted line) from
  \citet{Brandl04}. The spectrum of NGC\,7714 has been divided by 9.53
  so that its 14\,$\mu$m flux density matches that of SBS\,0335-052.
  Note the complete lack of PAHs in the spectrum of SBS\,0335-052 as
  well as the spectrum peak at $\sim$30\,$\mu$m (see Section 3.1).
  The solid circle is our 22\,$\mu$m peak-up photometric point, while
  the diamond corresponds to the 21\,$\mu$m Gemini point from
  \citet{Plante02}.}

\end{figure}

Our spectrum displays strong [SIV]$\lambda$10.51$\mu$m, and
[NeIII]$\lambda$15.55$\mu$m lines, while [SIII], [NeII], and H$_2$ 0-0
S(3) may also be present (see Table 1). The observed (not corrected
for extinction) [NeIII] and [SIV] fluxes in conjunction with the
free-free radio emission have been used to derive the ionic abundances
\citep{Osterbrock89}. This method is {\em a priori} more robust than
using hydrogen recombination lines because it is relatively free of
complicating factors: extinction, uncertainties about electron
temperatures and densities as well as underlying stellar absorption
and wind associated emission features.  Using the results of
\citet{Hunt04}, which suggest a free-free emission of
S$_{\nu}$=0.27\,mJy at 5\,GHz and electron density of
$\sim$2000\,cm$^{-3}$, as well as the helium abundance as given by
\citet{Izotov97} we find that Ne$^{++}$/H$^{+}$=1.05$\times10^{-5}$
which is $\sim$9\% of the solar value and
S$^{3+}$/H$^{+}$=4.3$\times10^{-7}$ or $\sim$2\% of solar
\citep{Anders89, Grevesse93}. These estimates are lower bounds on the
total Ne and S abundances because only a single ionic state has been
measured and extinction corrections have not been made. Had we
corrected for extinction based on the observed 9.7$\mu$m silicate
absorption, the corresponding A$_{10.51\mu \rm m}$= 0.33\,mag and
A$_{15.56\mu \rm m}$= 0.072\,mag would imply that the abundances could
be higher by a factor of 1.36 and 1.07 respectively. These abundances
are higher than the Z$_{\sun}$/41 metallicity that is derived for
oxygen from the optical observations and imply that the region probed
by our spectrum has polluted its environment by mass loss from
supernovae (SNe) and Wolf-Rayet stars.

However, there are a number of issues associated with the above
mentioned abundances which must be considered. There is an
inconsistency between the radio measurements and the fluxes of the
hydrogen recombination lines quoted in the literature. If we use the
5\,GHz free-free emission of \citet{Hunt04} to predict the
corresponding extinction corrected H$\beta$ flux we find a value of
6.1$\times10^{-14}$erg\,cm$^{-2}$\,s$^{-1}$ which is {\em lower} than
the observed (not corrected for extinction) value of
8.5$\times10^{-14}$erg\,cm$^{-2}$\,s$^{-1}$ by \citet{Izotov98}. The
predicted Br$\alpha$ flux is also a factor of $\sim$2--5 less than the
one observed and \citet{Hunt04} advocate that winds could account for
some of the Br$\alpha$ flux. It remains unclear though why these winds
would not affect the other recombination lines. Even if we were to
consider possible errors in the estimate of the thermal radio emission
the puzzle remains.  Clearly a fraction of the observed IR line fluxes
originates from the dust-free ``optical'' region of the galaxy.  To
estimate this we can use the optical [NeIII]$\lambda$3868\AA~
measurements and derived physical conditions of the interstellar gas
by \citet{Izotov97} to predict the corresponding
[NeIII]$\lambda$15.55$\mu$m flux. For a T$_e$=20,000\,K \citep[][
derives 19,200\,K for \ion{O}{3}]{Izotov97} we find that the
[NeIII]$\lambda$15.55$\mu$m flux predicted by the optical measurements
is 0.75$\times10^{-14}$erg\,cm$^{-2}$\,s$^{-1}$, $\sim$2 times lower
than what is observed. This would suggest that indeed there is a star
forming region of the galaxy which is not accessible in the optical.
The result though depends strongly on the electron temperature. If
T$_e$=15,000\,K then the predicted mid-IR flux is similar to the one
observed. However, this would place the Ne$^{++}$ region at the same
temperature as the O$^{+}$region; an unlikely situation. The 9.7$\mu$m
silicate feature clearly indicates that there is a powerful source of
luminosity hidden behind a screen of at least A$_{9.7\mu \rm m}\sim$
0.5\,mag or A$_{\rm V}\sim$15\,mag. Clearly all of these uncertainties
can strongly affect any abundance calculations.

\begin{deluxetable}{ccc}
\tabletypesize{\scriptsize}
\setlength{\tabcolsep}{0.02in}
\tablecaption{Mid-IR Line Fluxes of SBS\,0335-052\label{tbl1}}
\tablewidth{0pc}
\tablehead{ 
\colhead{Ion}  & \colhead{$\lambda (\mu m)$}  & \colhead{Flux ($\times10^{-17}$W m$^{-2}$)}
}
\startdata
$ \rm [SIV]$    & 10.51 & 1.62$\pm0.09$  \\
$ \rm [SIII]$   & 18.71 & $<$0.54 \tablenotemark{a} \\
$ \rm [NeII]$   & 12.81 & $<$0.28  \tablenotemark{a}\\
$ \rm [NeIII]$  & 15.55 & 1.40$\pm0.08$  \\
H$_2$ 0-0 S(3)  & 9.67  & $<$0.37 \tablenotemark{a}
\enddata

\tablenotetext{a}{The upper limits are 3-$\sigma$.}
\end{deluxetable}

Another major difference between the two spectra shown in Figure 1 is
that in SBS\,0335-052 the continuum shortward of $\sim$15\,$\mu$m
appears similar to that which underlies the emission features in
NGC\,7714.  However, these continua depart dramatically at longer
wavelengths.  The continuum of NGC\,7714 increases rapidly at longer
wavelengths because of a relatively more massive cool dust component
which characterizes many luminous infrared galaxies. Conversely, the
spectrum of SBS\,0335-052 peaks at 28\,$\mu$m.  In Figure 2 we have
drawn an offset power-law, $f_{\nu}\sim \nu ^{1.3}$, to the IRS
spectrum to extrapolate to longer wavelengths.  Using the non thermal
1.46GHz flux density of the galaxy measured by \citet{Hunt04} and the
radio to far-infrared correlation, we predict that the far-infrared
luminosity is F$_{fir}$(43--123\,$\mu$m)=2$\times10^{-15}$ Wm$^{-2}$.
The corresponding average flux density over this wavelength range is
$\sim$44\,mJy.  Even if the entire far-infrared luminosity originates
from the 60\,$\mu$m band, the corresponding flux density would be
58\,mJy, well below the \citet{Plante02} 112\,mJy measurement at
65\,$\mu$m. However, the 58\,mJy estimate is consistent with the IRS
spectrum which clearly decreases at wavelengths longer than
30\,$\mu$m.

\begin{figure}
\figurenum{2}
\epsscale{1.2}
\plotone{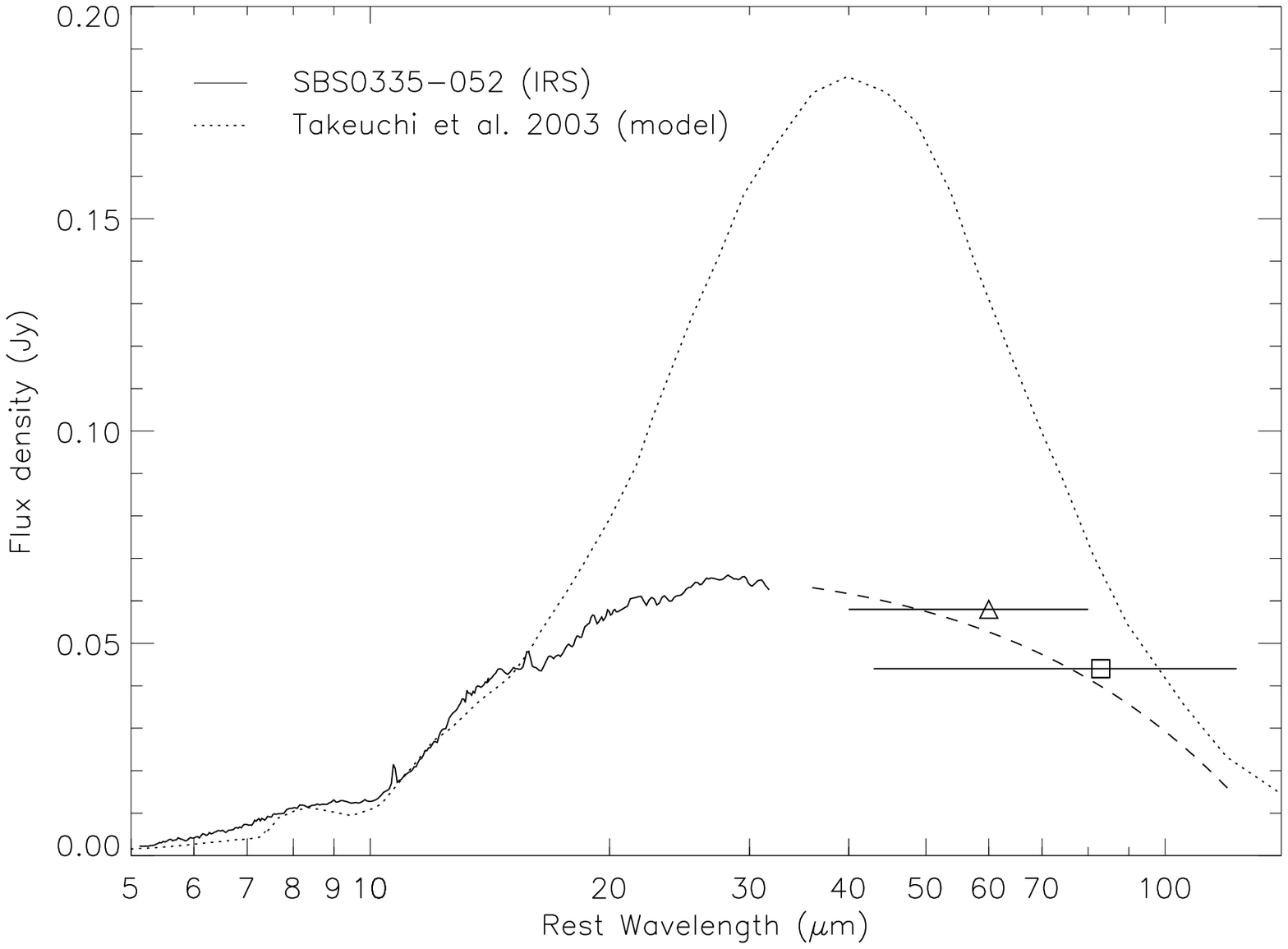}
 
\caption{
  The IRS spectrum of SBS\,0335-052 (solid line) along with the model
  of \citet{Takeuchi03} (dotted line). Note how the model deviates
  past 16\,$\mu$m. Based on the observed 5--35\,$\mu$m SED of the
  galaxy we can extrapolate to longer wavelengths (dashed line) using
  a an offset power-law peaking at 28\,$\mu$m with an asymptotic
  shape, $f_{\nu}(\nu \to 0) \sim \nu ^{1.3}$. The difference between
  the observed SED and the model has only a small effect in the
  infrared luminosity, but it substantially reduces the amount of cold
  dust in the galaxy.  We also show estimates derived from the radio
  data of the average flux density over the far-infrared range and the
  upper limit at 60\,$\mu$m (box and triangle, respectively; see
  Section 3).{\label{f2}} }
\end{figure}

These two major differences, no PAH features and relatively flat
continuum, provide an initial indication that the PAH criterion for
identifying starbursts may not apply to systems of low metallicity.
Mid-IR spectra of Active Galactic Nuclei (AGN) often display ionic
lines and also lack PAH emission features, but the continua are
typically even flatter between 5--20\,$\mu$m than SBS\,0335-052 (see
ISO observations of Seyfert 1s by Clavel et al. 2000, or the nuclear
spectrum of NGC\,1068 by Sturm et al. 2000 and Le Floc'h et al. 2001
as well as Peeters et al. 2004).  If metal poor galaxies such as
SBS\,0335-052 can be present at z$\sim$0, it is likely that similar
systems may exist not only as high-z primordial galaxies, but also at
intermediate redshifts $0.5<$z$<1.5$ where moderate luminosity
galaxies (L$_{\rm IR}<10^{11}$ L$_{\sun}$) are accessible with the
deep infrared surveys performed by Spitzer.  Far-infrared and
submillimeter photometric methods for estimating redshifts based on
spectral energy distributions (SEDs) of nearby metal rich systems,
with continua peaking at $\sim$100\,$\mu$m, would fail to identify
metal poor systems with SEDs similar to that of SBS\,0335-052, which
peaks at $\sim$28\,$\mu$m.

\subsection{IR luminosity and Dust Mass}

The luminosity of a dust obscured source is the integrated infrared
flux corrected for the object's distance.  However, determining the
total dust mass depends crucially on the dust's temperature,
distribution, and the optical properties.  \citet{Thuan99} concluded
from their 5--15\,$\mu$m ISO spectrum that the starburst responsible
for heating the dust which produces the mid-IR continuum in
SBS\,0335-052 is heavily obscured, with an optical extinction A$_{\rm
  V}$$\sim$20\,mag.  They also modeled the dust continuum as being
represented by a single temperature of about 260K and an emissivity,
$\epsilon\sim\nu^{1.5}$.  On this basis they suggested that the young
star clusters visible optically were independent of the obscured
clusters.  These general conclusions have been supported by
\citet{Hunt01,Hunt04}, \citet{Plante02}, and \citet{Takeuchi03}.  If
correct, this interpretation implies that primordial starbursts of
very low metallicity can nevertheless surround themselves with
sufficient dust to be completely hidden optically.  A very different
conclusion was reached by \citet{Dale01}, using ground-based mid-IR
imagery; they conclude that the optical extinction to the starburst
which heats the dust is less than one magnitude and that a two
component dust model is needed, with temperatures of about 80K and
210K. Both components are optically thin, and the bolometric
luminosity is dominated by the cooler component.

All previously published models extending beyond 20\,$\mu$m had their
assumed spectra anchored to the IRAS and ISO limits at 60 and
100\,$\mu$m, as well as to the ISO detection at 65\,$\mu$m
\citep{Dale01, Plante02}. The shape and intensity of the IRS spectrum
clearly demonstrates that those far-infrared values do not accurately
reflect the true shape of the spectrum, which is inconsistent with a
flux density above $\sim$70\,mJy anywhere in the 30--100\,$\mu$m range
(see Figure 2).  We have fit both the new IRS spectrum and the old ISO
spectrum with a simple two-temperature model with emissivity of
$\epsilon\sim\nu^{1.5}$.  In both cases good fits are obtained with
dust temperatures of $\sim$65 and $\sim$150K.  While the $\sim$150K
component is the same for both models, the required dust mass for the
cool component is 4 times less for the Spitzer spectrum.  Scaling to
the predicted 100\,$\mu$m flux, we estimate the mass of the cool dust
to be 6$\times10^3$ M$_{\sun}$ for the spectrum based on the ISO data
and 1.5$\times10^3$ M$_{\sun}$ for the IRS spectrum
\citep{Hildebrand83, Cox00}.  These masses are considerably below the
masses reported previously \citep[$\sim10^5$M$_{\sun}$,
see][]{Thuan99,Hunt01,Plante02}. Qualitatively this is because the
cold component in the Spitzer model is warmer and produces less flux
than the cold component in the ISO models. Dramatic changes in the
dust properties could increase the dust mass, but the materials used
in the cited models do not differ greatly from the standard.  Several
models, including the one developed by \citet{Takeuchi03}, derive a
large fraction of the dust components from the SNe ejecta produced in
the enshrouded super star clusters.  However, the amount of dust
produced by SNe in low metallicity systems is predicted to be rather
low \citep[i.e.][]{Todini01}.  Furthermore, in SNe remnants of our
Galaxy most of the dust appears to be very cold
\citep[T$\sim$18K,][]{Dunne03}.  The fact that the Spitzer data
suggest that much less cool dust is required eases the problem of
accounting for the {\em in situ} dust generation.

\section{Conclusions}

The very low metallicity Blue Compact Dwarf galaxy SBS\,0335-052 is
shown to have a very unusual spectrum which is quite different from
the spectrum of typical starburst galaxies --- the flux density,
f$_{\nu}$, peaks at $\sim$28\,$\mu$m while the luminosity,
$\nu$f$_{\nu}$, peaks at $\sim$20\,$\mu$m.  There are no detectable
PAH emission features. The spectrum is characterized by a warm
($\sim$150K) dust component with a more massive cool ($\sim$65K) dusty
envelope.  However, the mass of the cool region is far less than what
had been previously estimated. Silicate absorption features are
clearly present (A$_{9.7\mu \rm m} \geq 0.49$\,mag). Preliminary
analysis of the infrared ionic lines suggests that the young central
cluster may have already polluted the cocoon enshrouding it, and its
metallicity could be higher than that determined from optical
observations alone.  Upcoming mid-IR spectroscopy using the high
resolution modules of IRS will provide higher sensitivity and more
accurate line fluxes.

Theoretical models have indicated that the dust responsible for the
mid-IR emission can be explained by ejecta from type 2 SNe and
Wolf-Rayet stars in the buried super star cluster.  However, there are
no complete models to date that fit the mid-IR spectrum and include
dynamics of the dust generation and entrapment mechanisms.

\acknowledgments 

We would like to thank an anonymous referee whose comments greatly
improved the paper. The IRS was designed and built by Ball Aerospace
Corporation under contract from Cornell University.  Many dedicated
individuals at Ball made the IRS a success.  The IRS pipeline data
reduction tools were developed at the Spitzer Science Center.  This
work is based in part on observations made with the Spitzer Space
Telescope, which is operated by the Jet Propulsion Laboratory,
California Institute of Technology, under NASA contract 1407. Support
for this work was provided by NASA through Contract Number 1257184
issued by JPL/Caltech.

\end{document}